\documentstyle{elsart}
\newcommand{\gsim}{\mathrel{
\rlap{\lower4pt\hbox{\hskip1pt$\sim$}}
\raise1pt\hbox{$>$}}}
\newcommand{\lsim}{\mathrel{\rlap{\lower4pt\hbox{\hskip1pt$\sim$}}
\raise1pt\hbox{$<$}}}

\begin{document}
\begin{frontmatter}
\begin{flushright}
{\large\bf \hspace*{5cm} RESCEU-35/97\\
}
\end{flushright}
\title{First Results of Tokyo Dark Matter Search\\
with a Lithium Fluoride Bolometer}
\author[UT]{\underline{Wataru Ootani}\thanksref{e-mail}},
\author[UT]{Makoto Minowa},
\author[UT]{Takayuki Watanabe},
\author[UT]{Yutaka Ito\thanksref{ito}},
\author[UT]{Yasuhiro Kishimoto\thanksref{kishimoto}},
\author[UT]{Kentaro Miuchi},
\author[ICEPP]{Yoshizumi Inoue} and 
\author[CRC]{Youiti Ootuka}
\address[UT]{Department of Physics, School of Science, University of
Tokyo, 7-3-1 Hongo, Bunkyo-ku, Tokyo 113, Japan}
\address[ICEPP]{International Center for Elementary Particle Physics, 7-3-1
Hongo, Bunkyo-ku, Tokyo 113, Japan}
\address[CRC]{Cryogenic Center, University of Tokyo, 2-11-16 Yayoi,
Bunkyo-ku, Tokyo 113, Japan}
\thanks[e-mail]{E-mail address: wataru@icepp.s.u-tokyo.ac.jp}
\thanks[ito]{Present address: KEK, High Energy Accelerator Research
Organization, 3-2-1 Midori-cho, Tanashi-shi, Tokyo 188, Japan}
\thanks[kishimoto]{Present address: Institute for Chemical Research,
Kyoto University, Uji, Kyoto 611, Japan}
\begin{abstract}
The First results of the Tokyo dark matter search programme using a
21-g lithium fluoride bolometer are presented. The background spectrum was
measured in the surface laboratory. We derive an exclusion plot for
the spin-dependently coupled
Weakly Interacting Massive Particles (WIMPs) cross section.
\end{abstract}
\end{frontmatter}
\newpage
\section{Introduction}
The Weakly Interacting Massive Particles (WIMPs) are promising
candidates for the dark matter which might comprise a large fraction
of the mass of the Universe.
A number of groups are developing low background detectors which have enough 
sensitivity to detect low energy nuclear recoils caused by
the elastic scatterings of the WIMPs off nuclei.
Theoretical calculations of the cross sections of the elastic
 scattering for various detector materials have been worked out by
 many authors.
The cross section is usually separated into two independent parts;
spin-independent and spin-dependent part.
The spin-independent cross section is essentially proportional to the square
of the number
of nucleons in the nuclei of the detector material. 
Therefore, heavier nuclei are
suited for the spin-independently interacting WIMPs.
On the other hand,
the spin-dependent cross section is a function of nuclear spin and the 
Land\'e factor of the nucleus used in the detector, where the
 Land\'e factor measures
the spin factor of an unpaired nucleon in the nucleus.  The cross section
 is also a function of the quark spin contents of the unpaired nucleon.
Comparing these factors for various nuclei, one finds $^{19}$F to be the
best material to detect the spin-dependently interacting WIMPs\cite{Ellis}. 
For that reason, we have been working to develop 
the LiF bolometer\cite{Tokyo}. 
Prior to the planned underground experiments, the preliminary
measurements were performed with a 21-g LiF bolometer at the surface
laboratory. The results presented here are the first from the Tokyo dark
matter search programme.

\section{Experimental set-up and measurements}
The schematic view of the bolometer is shown in Fig.\ \ref{fig:bolometer}.
The bolometer is mounted in a home-made dilution refrigerator placed
at the surface laboratory of the University of Tokyo.
The 21-g LiF crystal ($20\times 20\times 20\, {\rm mm}^3$) is
thermally anchored by two copper ribbons to the
mixing chamber of the refrigerator which is cooled down to 10\,mK.
The temperature of the mixing chamber is monitored by
a cerous magnesium nitrate (CMN) magnetic susceptibility thermometer.
A home-made high-sensitivity neutron transmutation doped (NTD) 
germanium thermistor\cite{NTD} ($1.5\times 1\times 1\, {\rm mm}^3$) 
is attached to the crystal with GE varnish. 
The temperature of the crystal is not measured directly, but the zero
bias resistance of the thermistor implies that it is a little higher than
the temperature of the mixing chamber.

The radioactivity of this crystal was checked by a 
low-background Ge gamma-ray spectrometer before constructing the bolometer. 
The concentration
of radio-contaminations was less than 0.2\,ppb for U, 1\,ppb for Th,
and 2\,ppm for K.
The refrigerator is mostly made of low-radioactivity materials, which were also checked by the Ge gamma-ray spectrometer. 
The concentration of radio-contaminations of the refrigerator
materials were less
than 8\,ppb for U, 20\,ppb for Th, and 10\,ppm for K.
In particular, the refrigerator materials near the detector crystal
were selected
carefully. No detectable radioactivity was observed in the radio-assay
with the Ge gamma-ray spectrometer. 
The refrigerator is surrounded by lead shield 
with a thickness of 10\,cm. 

A voltage change across the dc-biased NTD Ge thermistor is fed into the 
source follower placed at the stage with a temperature of 4\,K, which
includes a low-noise J-FET heated up to 110\,K.
The signal is then fed into an external low-noise amplifier outside 
the refrigerator and the pulse shape of the signal is recorded by a
digital oscilloscope for off-line analysis.

The detector is calibrated using gamma-ray sources ($^{60}$Co, $^{137}$Cs, and $^{241}$Am) and an alpha-ray source ($^{241}$Am). 
The $^{241}$Am source was set
inside the cryostat in the calibration run previous to the
background measurements. In the periodic calibration during the
background measurements, the $^{60}$Co and $^{137}$Cs sources are set
between the cryostat and the lead shield.
Compton edges are used for the calibration by high energy gamma rays
which do not make enough photoelectric peak in the low atomic number
LiF crystal.
An energy resolution for the 60\,keV gamma-ray from $^{241}$Am is
4.8\,keV (FWHM), which is limited predominantly by the base-line
fluctuation. Good linearity is confirmed up to about 5\,MeV as shown
in Fig.\ref{fig:linearity}, where the detector gain for alpha-rays is
the same as that for gamma-rays. It implies that the
bolometer has no suppression on the collection of
the nuclear recoil energy. It must be noted that for the detectors based on
ionization such as semiconductor detector and scintillator, the
observed energy is generally less than the recoil energy.
The detector gain is fairly stable within $\pm$5\%
against the change of the temperature of the mixing chamber from 9 mK to 13 mK.
This feature is desirable for the long term running to search for the WIMPs.

\section{Results and discussions}
With this 21-g LiF bolometer, we made a test running at the surface
laboratory at the Hongo campus of the University of Tokyo.
Fig.\ \ref{fig:bg spectrum}
 shows the background spectrum corresponding to an exposure of
$21\,{\rm g}\times1.19\,{\rm days}$.
Relatively flat continuum down to 14\,keV is observed and 
the counting rate increases steeply below 14\,keV. 
This increase is mainly due to pick-up noise. 
The energy spectrum above 14\,keV, therefore, is used in our analysis. 
The counting rate
itself is still too high because of the background radiation in the
surface laboratory and incompleteness of the radiation shield. 

We evaluate an exclusion plot for
the spin-dependently interacting WIMP cross section from the
background data obtained by this short test running.
The lower limit of the WIMP-nucleus cross section, $\sigma_{{\rm
WIMP}-N}$, at 95\% CL for a fixed WIMP mass 
can be derived by comparing 
the obtained background spectrum with the theoretical one.
This comparison is done in each energy bin above 14\,keV and 
the lowest value of the cross-section is adopted.
The theoretical recoil spectrum for a given WIMP mass is calculated 
assuming a Maxwellian dark matter velocity distribution with rms
velocity of 230\,km/s, and 
then folded with the measured
energy resolution and the nuclear form factor. 
We also assume the local density of the WIMP to be 0.3\,GeV/cm$^3$. 

It is customary to convert thus obtained $\sigma_{{\rm WIMP}-N}$ into a 
point-like cross section incident on a single proton, $\sigma_{{\rm
WIMP}-p}$, in order to compare the
results of various experiments with different nuclei with one another.
We follow the formula (6.6) in Ref.\ \cite{smith} 
with the spin factor for the fluorine of the odd group model given in  
Ref.\ \cite{Ellis}.
In Fig.\ \ref{fig:exclusion plots}, our 95\% CL exclusion limits on
$\sigma_{{\rm WIMP}-p}$ are 
plotted against the WIMP mass.

We also calculated the exclusion limits
for the results of BPRS
collaboration\cite{BPRS}, UKDMC\cite{UKDMC}, and
EDELWEISS collaboration\cite{EDELWEISS} in the same manner
as used in our analysis.  They are shown in the same figure for
comparison.  These three experiments were performed in deep
underground laboratories over a long term. 
Our short-term and small-size experiment in
the surface laboratory enabled us to place the exclusion
limit which is only a factor of three and ten behind that of EDELWEISS
and BPRS group respectively at the WIMP mass of 20\,GeV.
It demonstrates the advantages of the use of $^{19}$F and the 
low energy threshold of the bolometer for nuclear recoil.

\section{Prospects}
We have already constructed a multiple array of eight pieces of 21-g LiF
bolometers for the underground experiments and confirmed that 
each detector has similar performance to that of the single 21-g LiF 
bolometer used in this work. Installation of the arrayed bolometer in
the Nokogiri-yama
underground laboratory of the Institute for Cosmic Ray Research,
University of Tokyo is in progress.
The laboratory is 180\,m w.e. deep 
and located about 60\,km to the south of Tokyo.
We are going to make engineering runs to make
experiences to operate the cryogenic detectors at an underground
laboratory. 
The arrayed bolometer will be shielded with 10\,cm-thick
oxygen-free copper, 15\,cm-thick lead, 1\,g/cm$^2$-thick boric acid, 
and 20\,cm-thick polyethylene.  
Furthermore, a muon veto system with 2\,cm-thick plastic scintillator
will be employed in order to reduce the background produced by the
cosmic ray muon. One of the primary background sources in an
underground WIMPs search is fast neutron which produces the
nuclear recoil.
The ambient fast neutron flux at the Nokogiri-yama
underground laboratory is $2.3\times10^{-4}$\,$n$/cm$^2$/s which is 1/50 of that at
the surface laboratory of the University of Tokyo and expected to  be reduced with polyethylene shield by a factor of five.
The measurements in the Nokogiri-yama underground laboratory is expected
to bring the exclusion limit on $\sigma_{{\rm WIMP}-p}$ below the
limit of UKDMC.
The measurements in a still deeper underground site will enable us to
reach the sensitivity of $\sigma_{{\rm WIMP}-p}\sim0.01\ {\rm pb}$,
which is comparable to the predicted SUSY WIMP (neutralino) cross section.

\section*{Acknowledgements}
This work is supported by
the Grant-in-Aid in Scientific Research (A) and the Grant-in-Aid 
for COE Research by the Japanese Ministry of
Education, Science, Sports and Culture.

\newpage
\begin{figure}
\caption{Schematic drawing of the LiF bolometer.}
\label{fig:bolometer}
\end{figure}
\begin{figure}
\caption{Measured linearity of the bolometer. The solid line is the best
fit to the data.}
\label{fig:linearity}
\end{figure}
\begin{figure}
\caption{Background spectrum obtained with the LiF
bolometer. An exposure is 21\,g$\times$1.19\,days.}
\label{fig:bg spectrum}
\end{figure}
\begin{figure}
\caption{Exclusion plot derived from the background spectrum shown in
Fig.\ \protect\ref{fig:bg spectrum} for spin-dependent interacting WIMPs as a
function of the WIMP mass. For comparison, exclusion plots derived from 
the data in Ref. \protect\cite{BPRS,UKDMC,EDELWEISS} and scatter plot 
predicted in MSSM (10\,GeV$\leq
M_2\leq$1\,TeV, 10\,GeV$\leq \left|\mu\right|\leq$1\,TeV, $\tan\beta=2,8$) are also given. }
\label{fig:exclusion plots}
\end{figure}
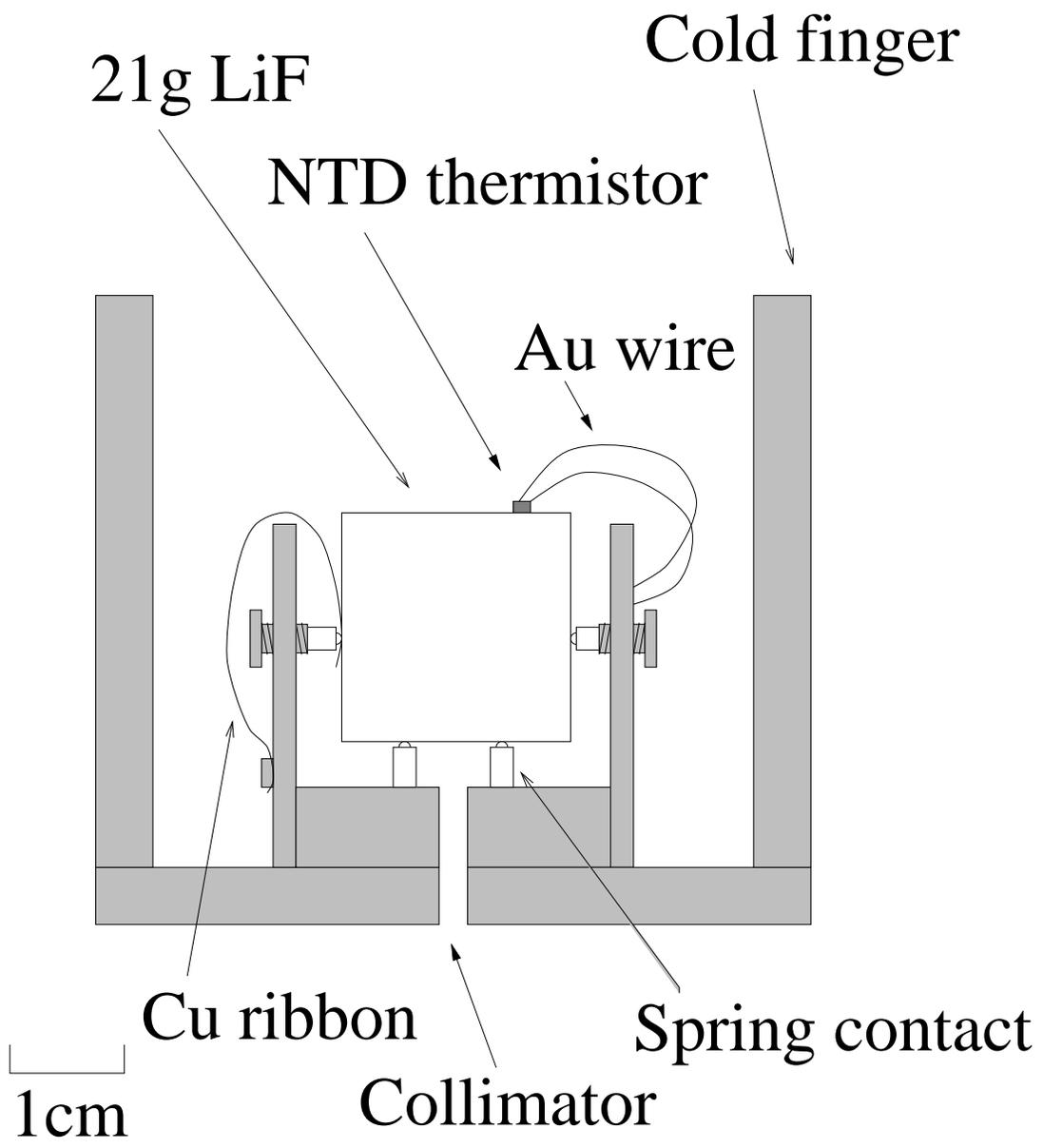
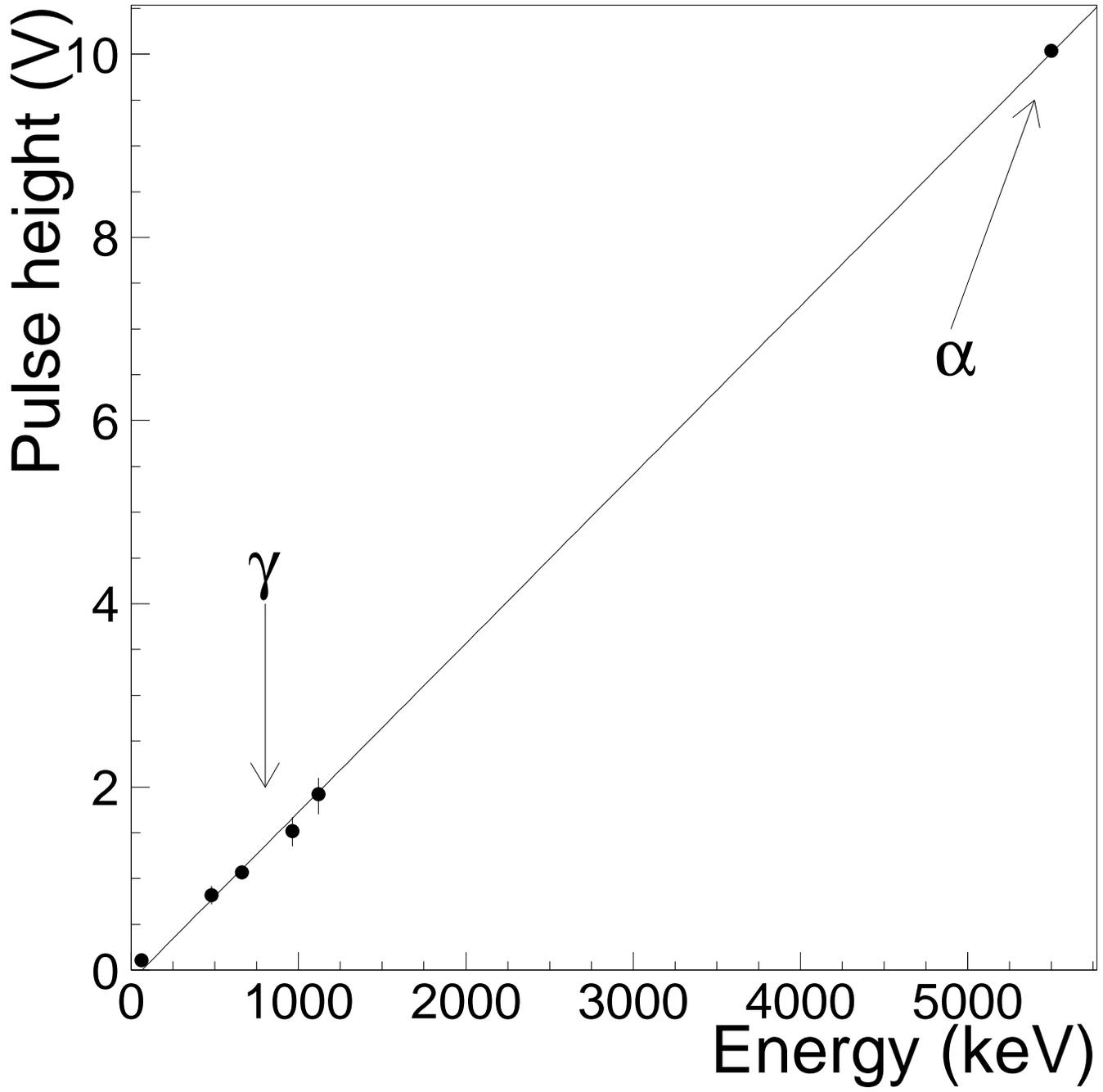
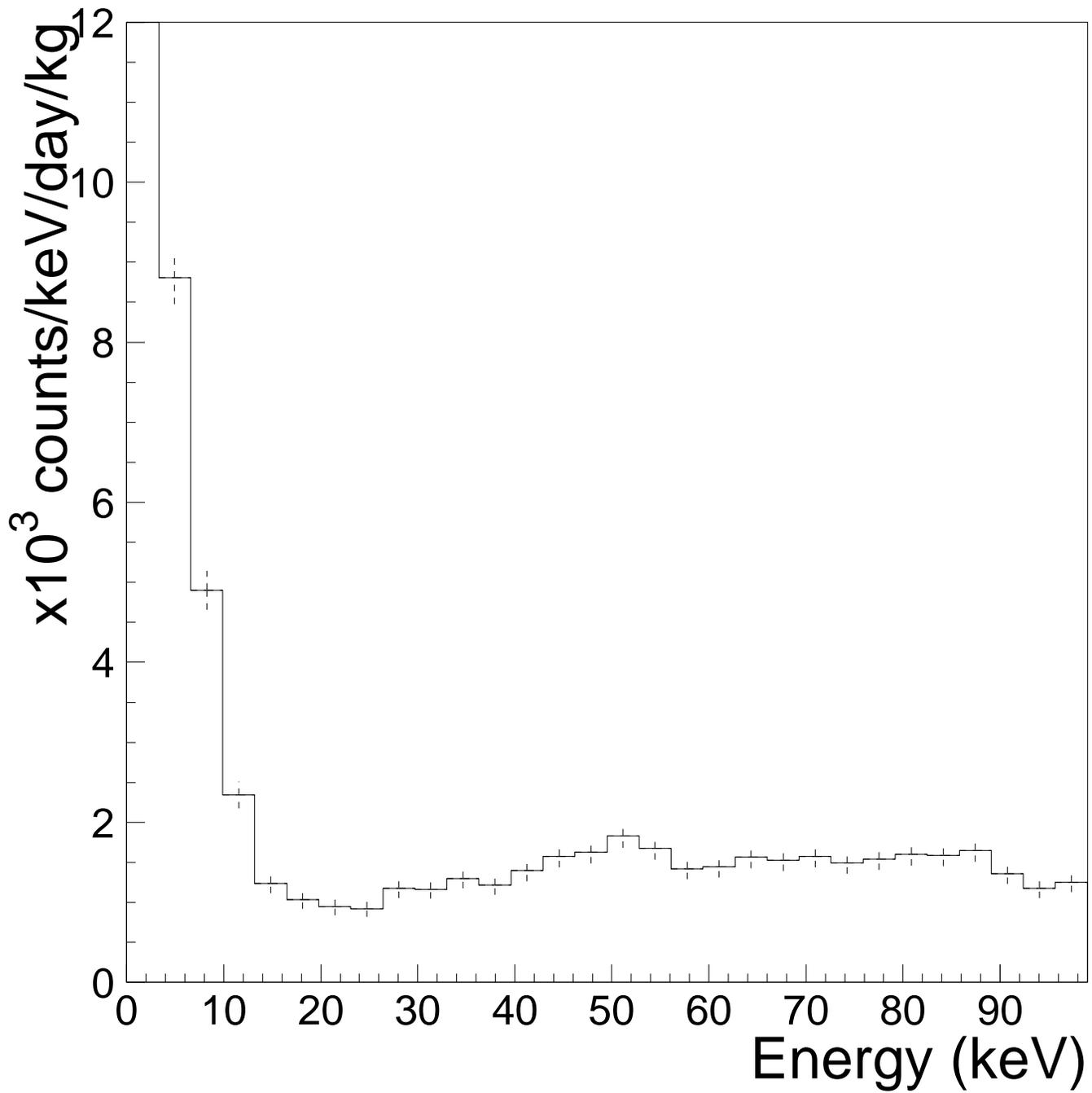
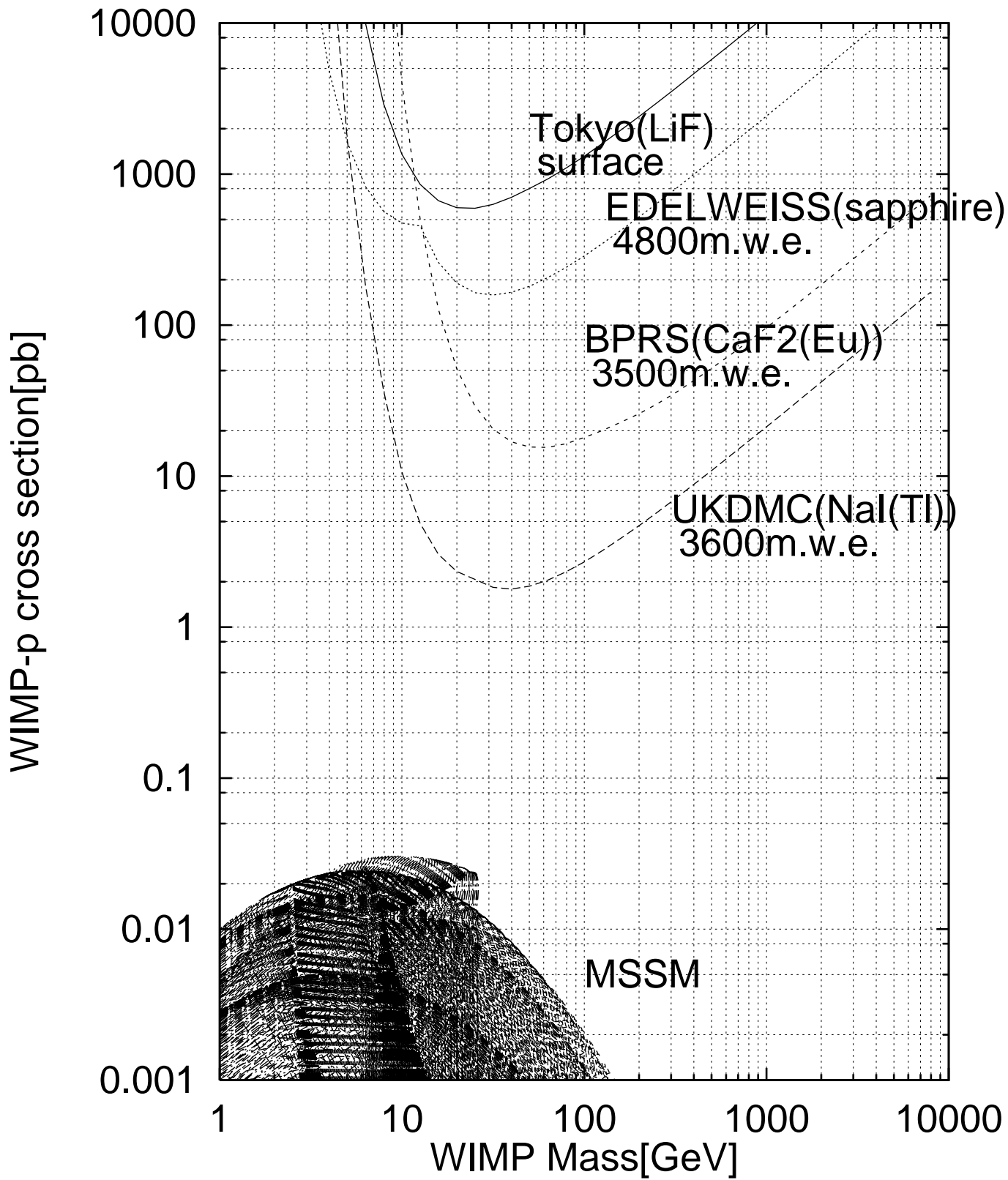
\end{document}